\documentclass{article}
\usepackage{graphicx}
\usepackage{amssymb}
\usepackage{amsmath}
\usepackage{times}
\usepackage[]{eufrak}
\usepackage{color}

\begin{document}
\begin{center}
{\bf\Large{ Quantum Interference Effects in Spacetime of Slowly
Rotating Compact Objects in Braneworld}}
\end{center}

\begin{center}
{Mamadjanov A.I.\footnote{Ulugh Beg Astronomical Institute
Astronomicheskaya 33,  Tashkent 100052, Uzbekistan}, Hakimov
A.A.\footnote{Institute  of Nuclear Physics, Tashkent 100214,
Uzbekistan}, Tojiev S.R.\footnote{National University of
Uzbekistan, Tashkent 100095, Uzbekistan}\\}
%
%{Institute of Nuclear Physics, Tashkent 100214, Uzbekistan\\    Ulugh Beg Astronomical Institute    Astronomicheskaya 33,    Tashkent 100052, Uzbekistan\\ National University Uzbekistan, Tashkent 100095, Uzbekistan }
\end{center}

\normalsize {\bf{Abstract}}

The phase shift a neutron interferometer caused by the
gravitational field and the rotation of the earth is derived in a
unified way from the standpoint of general relativity. General
relativistic quantum interference effects in the slowly rotating
braneworld as the Sagnac effect and phase shift effect of
interfering particle in neutron interferometer are considered. It
was found that in the case of the Sagnac effect the influence of
brane parameter is becoming important due to the fact that the
angular velocity of the locally non rotating observer must be
larger than one in the Kerr space-time. In the case of neutron
interferometry it is found that due to the presence of the
parameter $Q^{*}$ an additional term in the phase shift of
interfering particle emerges from the results of the recent
experiments we have obtained upper limit for the tidal charge as
$Q^{*}\lesssim 10^{7} \rm{cm}^{2}$.  Finally, as an example, we
apply the obtained results to the calculation of the (ultra-cold
neutrons) energy level modification in the braneworld.

\section{Introduction}
\normalsize  The detection of gravitational radiation provides not
only a verification of the predictions of general relativity, but
also opens a new window for astronomical observations. In this
respect, it is desirable to understand thoroughly the general
relativistic effects on the quantum mechanical interference. It
would be of great significance to have a formalism to derive these
various effects in a unified way. The aim of this paper is to show
such a treatment. We show that the Sagnac effect and phase shift
effect in a neutron interferometer in the braneworld.

The idea that our Universe might be a three-brane \cite{randal},
emdedded in a higher dimensional spacetime, has recently attracted
much attention.  Static and spherically symmetric exterior vacuum
solutions of the brane world models were initially proposed by
Dadhich et al \cite{dadhich,maartens} which have the mathematical
form of the Reissner-Nordstr\"{o}m solution, in which a tidal Weyl
parameter $Q^*$ plays the role of the electric charge of the
general relativistic solution. The role of the tidal charge in the
orbital resonance model of quasiperiodic oscillations in black
hole systems \cite{stuklik} and in neutron star binary systems
\cite{stuklik2} have been studied intensively. The so-called DMPR
solution was obtained by imposing the null energy condition on the
three-brane for a bulk having nonzero Weyl curvature. In this
paper we testing the Sagnac effect and phase shift effect in a
neutron interferometer in that the braneworld.

The experiment to test the effect of the gravitational field of
the Earth on the phase shift in a neutron interferometer were
first proposed by Overhauser and Colella \cite{overhauser}. Then
this experiment was successfully performed by Collela, Overhauser
and Werner \cite{colella}. After that, there were found other
effects, related with the phase shift of interfering particles.
Among them the effect due to the rotation of the Earth
\cite{page,werner}, which is the quantum mechanical analog of the
Sagnac effect, and the Lense-Thirring effect \cite{mashhoon} which
is a general relativistic effect due to the dragging of the
reference frames. So we do not consider the neutron spin in this
paper

In the paper \cite{kuroiwa} a unified way of study of the effects
of phase shift in neutron interferometer due to the various
phenomena was proposed. Here we extend this formalism to the case
of the slowly rotating braneworld in order to derive such phase
shift due to either existence or nonexistence of the tidal brane
charge.

The Sagnac effect is well known and thoroughly studied in the paper \cite{G.rizzi}. It
presents the fact that between light or matter beam
counter-propagating along a closed path in a rotating
interferometer a fringe shift $\triangle \varphi$ arises. This
phase shift can be interpreted as a time delay $\triangle T$
between two beams, as it can be seen below, doesn't include the
mass or energy of particles. That is why we may consider the
Sagnac effect as the "universal" effect of the geometry of
space-time, independent of the physical nature of the interfering
beams. Here we extend the recent results obtained in the papers
\cite{rizzi,ruggiero} where it has been shown a way of calculation
of this effect in analogy with the Aharonov-Bohm effect, to the
case of slowly rotating braneworld.

The paper is organized as follows. In section \ref{2ndsec}, we
taking starting from the covariant. Klein-Gordon equation in the
braneworld , and consider terms  the phase difference of the wave
function. In section \ref{3rdsec} we consider interference in
Mach-Zender interferometer in the background of spacetime of black
hole in braneworld. Section \ref{4thsec} is devoted to study
Sagnac effect in the slowly rotating braneworld.

Recently Granit experiment \cite{nesvizhevsky} verified the quantization
of the energy level of ultra-cold neutrons (UCN) in the Earth's
gravity field and new, more precise experiments are planned to be performed.
Experiments with UCN have high accuracy and that is the reason to look for
verification of the gravitomagnetism in such experiments. As an example we
investigate modification of UCN energy levels caused by the existence of brane parameter.

Throughout, we use space-like signature ( -, +, +, +) (However,
for those expressions with an astrophysical application we have
written the speed of light explicitly). Greek indices are taken to
run from 0 to 3 and Latin indices from 1 to 3; covariant
derivatives are denoted with a semi-colon and partial derivatives
with a comma.

\section{The Phase shift}\label{2ndsec}

\normalsize

We assume that the external gravitational field of the earth is
described by the braneworld model proposed recently in papers
\cite{kuroiwa}. Slowly rotating gravitating object in braneworld
in a spherical coordinate system is described by the metric .

\begin{equation}\label{metrik}
ds^{2}=-A^{2}dt^{2}+H^{2}d
r^2+r^{2}d\theta^{2}+r^{2}\sin^{2}\theta
d\phi^{2}-2\widetilde{\omega}(r)r^{2}\sin^{2}\theta dtd \phi ,
\end{equation}
here
\begin{equation}
A^{2}(r)\equiv\left(1-\frac{2M}{r}+\frac{Q^{*}}{r^{2}}\right)=H^{-2}
,
\end{equation}
is the Reissner-Nordstr\"{o}m-type exact solution \cite{dadhich}
for the metric outside the gravitating object and $
\widetilde{\omega}(r)=\omega(1-\frac{Q^{*}}{2rM})=2Ma/r^{3}(1-Q^{*}/2rM)$.

Using the Klein-Gordon equation
\begin{eqnarray}
&& \nabla^{\mu}\nabla_{\mu}\Phi-(mc/ \hbar)^{2}\Phi=0 ,
\end{eqnarray}
for particles which have a mass $m$, in the paper \cite{kuroiwa}
it was defined the wave function $\Phi$ of interfering particles
as
\begin{eqnarray}
&& \Phi=\Psi exp\left(-i\frac{mc^2}{\hbar}t\right) \ ,
\end{eqnarray}
where $\Psi$ is the nonrelativistic wave function.

In the present situation, both $GM/rc^{2}$ and $a/r$ are
sufficiently small and their higher order terms can be neglected.
Therefore, to the first order in $M$ and $Q$ and neglecting the
terms of $O((\upsilon/c)^{2})$, the Klein-Gordon equation in the
braneworld becomes
\begin{equation}
\label{Schr}
 i\hbar\frac{\partial \Psi}{\partial
t}=-\frac{\hbar^2}{2m}\left[\frac{1}{r^2}\frac{\partial}{\partial
r}\left(r^2\frac{\partial}{\partial
r}\right)-\frac{L^2}{r^2\hbar^2}\right]\Psi-\frac{GM
m}{r}\Psi+\frac{Q^{*}mc^{2}}{2r^{2}}\Psi
 +\frac{2GMa}{r^{3}c}(1-\frac{Q^{*}}{r^{2}}-\frac{Q^{*}c^2}{2rMG})L_z\Psi  \ ,
\end{equation}
where we have used the following notations:
\begin{eqnarray}
&& L^{2}=-\hbar^{2}\left[\frac{1}{\sin\theta}\frac{\partial}
{\partial\theta}\left(\sin\theta\frac{\partial}{\partial\theta}\right)
+\frac{1}{\sin^{2}\theta}\frac{\partial^{2}}{\partial\phi^{2}}
\right] \ ,
\\
&& L_z=-i\hbar\frac{\partial}{\partial\phi}\ ,
\end{eqnarray}
which correspond the square of the total orbital angular momentum
and $z$ component of the  orbital angular momentum operators  of
the particle with respect to the center of the earth,
respectively. After the coordinate transformation
$\phi\rightarrow\phi+\Omega t$, where $\Omega$ is the angular
velocity of the earth, we obtain the Schr\"{o}dinger equation for
an observer fixed on the earth in the following form:
\begin{eqnarray}
\label{Schr} &&i\hbar\frac{\partial \Psi}{\partial
t}=-\frac{\hbar^2}{2m}\left[\frac{1}{r^2}\frac{\partial}{\partial
r}\left(r^2\frac{\partial}{\partial
r}\right)-\frac{L^2}{r^2\hbar^2}\right]\Psi-\frac{GM
m}{r}\Psi+\frac{Q^{*}mc^{2}}{2r^{2}}\Psi-\Omega L_{z}\Psi\nonumber\\
&&
\qquad\qquad+\frac{2GMa}{r^{3}c}L_z\Psi-\frac{GQ^*ac}{r^4}\left(1+\frac{2
GM}{c^2r}\right)L_z\Psi \ .
\end{eqnarray}
The Hamiltonian derived in the last section can be divided in to
the five terms as:
\begin{eqnarray}
H=H_{0}+H_{1}+H_{2}+H_{3}+H_{4} ,
\end{eqnarray}
where
\begin{eqnarray}
&&H_0=-\frac{\hbar^2}{2m}\frac{1}{r^2}\frac{\partial}{\partial
r}\left(r^2\frac{\partial}{\partial r}\right)+\frac{L^2}{2mr^2}\
 ,\quad H_1=-\frac{GM m}{r}+\frac{Q^{*}mc^{2}}{2r^{2}}\ ,  \quad H_2=-\Omega L_z\ , \quad\nonumber\\
&&H_3=\frac{2GM a}{r^3c}L_z\ , \quad \quad
H_{4}=-\frac{Q^*ac}{r^4}\left(1+\frac{2GM}{c^2r}\right)L_z .
\label{hamiltonian}
\end{eqnarray}
 $H_{0}$ is the Hamiltonian for a freely propagating particle,
$H_{1}$ is the Newtonian gravitational potential energy, $H_{2}$
is concerned to the rotation of the gravitating source, $H_{3}$ is
related to the Lense-Thirring effect (dragging of the inertial
frames). The phase shift terms due to $H_1, H_2$ and $H_3$ are
\begin{eqnarray}
&&\beta_{Sag}\simeq\frac{2m \bf{\Omega} \cdot\textbf{S}}{\hbar}\ ,
\quad \beta_{drag}\simeq\frac{2Gm}{\hbar
c^2R^3}\textbf{J}\cdot\left[\textbf{S}-3\left(\frac{\textbf{R}}{R}\cdot\textbf{S}\right)
\frac{\textbf{R}}{R}\right] \ ,
\end{eqnarray}
respectively and in the first place, let us calculate the phase
shift due to the gravitational potential. For the purpose of the
present discussion, the quasi-classical approximation is valid and
the phase shift is given by the integration along a classical
trajectory (see.fig 1)
\begin{eqnarray}
\beta_{grav}=\beta_{ABD}-\beta_{ACD}=-\frac{1}{\hbar}\int\frac{GQ^*a}{r^4}\left(1+\frac{2M}{r}
\right)d
r\simeq\frac{m^2S\lambda}{2\pi\hbar^2}\left(g-\frac{Q^*c^2}{R^3}
\right)\sin\phi\ \ .
\end{eqnarray}
 The phase difference $\beta_{grav}$ will be
\begin{equation}
\beta_{grav}=\left[\frac{m^{2}S \lambda g}{2\pi
\hbar^{2}}-\frac{m^{2}S \lambda}{2\pi
\hbar^{2}}\cdot\frac{Q^{*}c^2}{R^3}\right]=\beta\pm\Delta\beta \ .
\end{equation}
\begin{center}
\begin{figure}
\includegraphics[width=0.75\textwidth]{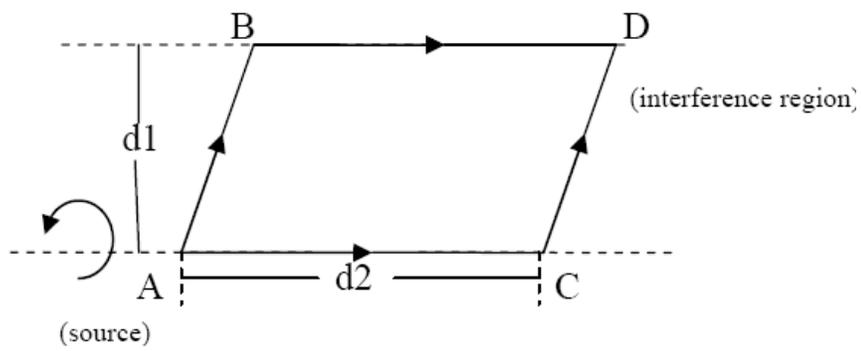}
\caption{Schematic illustration of alternate paths separated in
vertical direction in a neutron interferometer.} \label{fig1}
\end{figure}
\end{center}

%\begin{center}
%\begin{figure}
 % \includegraphics[width=0.75\textwidth]{Mach-Zender_interforometry.eps}
%\caption{{The gradient of $U$ in the interferometer's rest frame.
%$h$ and $l$ are the interferometer's height and length.}}
%\end{figure}
%\end{center}

%
As we know according to the experiment \cite{colella} the phase
difference due to the gravitational potential of the earth was in
good agreement with the theoretical prediction within an error of
$1 \%$.
Therefore one can easily obtain the following upper limit for the
module of the tidal brane charge as:

\begin{equation}
Q^*
%=\frac{\eta g R^3}{c^2}=3\cdot10^{6}sm^2<
\lesssim10^{7} \rm{sm}^2 \ .
\end{equation}
 Here $ S=d_1d_2$ is the area of interferometer, $\bf S$
is the area vector of the sector ABCD (see fig. 1),
${\bf\Omega}=(0,0,\Omega)$ and ${\bf J}=(0,0,J)$ are the angular
velocity and the angular momentum vectors of the object
correspondingly, $\bf R$ is the position vector of the instrument
from the center of the gravitating object, $\lambda$ is de Broglie
wavelength.

The last term of the equation (\ref{hamiltonian}) represents the
parts $H_4$ Hamiltonian

\begin{eqnarray}
H_{4}=-\frac{Q^*ac}{r^4}\left(1+\frac{2GM}{c^2r}\right)L_z \ ,
\end{eqnarray}
related to the $Q^*$-Brane parameter.

Integrating it over time along the trajectory of the particle, one
can find the corresponding phase shift
\begin{eqnarray}
\beta_4=-\frac{1}{\hbar}\int\frac{2GMQ^{*}a}{r^5c}L_z dt,  \quad
\beta_5=-\frac{1}{\hbar}\int\frac{Q^{*}ac}{r^4}L_z dt \ .
\end{eqnarray}

Presenting ${\bf r}={\bf R}+{\bf r'}$, where $\bf r'$ denotes
position of the given point of the interferometer from the center
of the instrument, and assuming that $r'/R$ is small one can
obtain that the angular dependence of $\beta_{brane}$
\begin{eqnarray}
&&\beta_{brane}=\beta_{4(ABD)}-\beta_{4(ACD)}=\frac{2GQ^*m}{\hbar
c^2}\oint\frac{{\bf J}\cdot({\bf r}\times d{\bf
r})}{r^5}=\frac{2GQ^*m}{\hbar c^2}{\bf J}\cdot\oint\frac{({\bf
R}+{\bf r'})\times d{\bf r'}}{|{{\bf
R}}+{\bf r'}|^5}\nonumber\\
&&\simeq-\frac{2GQ^*m}{\hbar c^2R^5}{{\bf J}}\cdot\left[\oint {\bf
r'}\times d{\bf r'}-5\oint\left(\frac{{\bf R}}{R}\cdot
\textbf{r}'\right)\frac{{{\bf R}}}{R}\times
d\textbf{r}'\right]=-\frac{2GQ^*m}{\hbar c^2R^5}{{\bf
J}}\left[\textbf{S}-5\left(\frac{{\bf R}}{R}\cdot
\textbf{S}\right)\frac{{\bf R}}{R}\right] \ ,
\end{eqnarray}
and
\begin{eqnarray}
&&\beta'_{brane}=\beta_{5(ABD)}-\beta_{5(ACD)}=\frac{Q^*m}{\hbar M
}\oint\frac{{\bf J}\cdot(\textbf{r}\times
d\textbf{r})}{r^4}=\frac{Q^*m}{\hbar M }{\bf
J}\cdot\oint\frac{({\bf R}+\textbf{r}')\times d\textbf{r}'}{|{\bf
R}+\textbf{r}'|^4}\nonumber\\
&&\simeq-\frac{Q^*m}{\hbar M R^4}{\bf J}\cdot\left[\oint
\textbf{r}'\times d\textbf{r}'-4\oint\left(\frac{{\bf R}}{R}\cdot
\textbf{r}'\right)\frac{{\bf R}}{R}\times
d\textbf{r}'\right]=-\frac{Q^*m}{\hbar MR^4}{\bf
J}\left[\textbf{S}-4\left(\frac{{\bf R}}{R}\cdot
\textbf{S}\right)\frac{{\bf R}}{R}\right]\ ,
\end{eqnarray}
where ${\bf R}$ represents the position  vector of the instrument
from the center of the earth. If we assume that the earth is a
sphere of radius $R$ with uniform density, then
\begin{eqnarray}
{\bf J}=\frac{2}{5}M R^2{\bf \Omega} \ ,
\end{eqnarray}
and, if {\bf R} is perpendicular to $\bf S$
\begin{eqnarray}
\beta_{brane}=-\frac{1}{5}\frac{r_g Q^*}{R^3}\beta_{Sag}, \quad
\beta'_{brane}=-\frac{1}{5}\frac{Q^*}{R^2}\beta_{Sag}\ ,
\end{eqnarray}
if {\bf R} is parallel to $\bf S$
\begin{eqnarray}
\beta_{brane}=\frac{4}{5}\frac{r_g Q^*}{R^3}\beta_{Sag}, \quad
\beta'_{brane}=\frac{3}{5}\frac{Q^*}{R^2}\beta_{Sag} \ ,
\end{eqnarray}
where $r_g=2GM/c^2$ is the Schwarzchild radius of earth.

%
%  Ultra cold neutrons
%
Astrophysically it is interesting to apply the obtained result for
the Hamiltonian of the particle, moving around rotating
gravitating object in braneworld, to the calculation of energy
level of ultra-cold neutrons (UCN) (as it was done for slowly
rotating Kerr space-time in the work \cite{arminjon} and for
slowly rotating object with nonvanishing gravitomagnetic charge in
the work \cite{vika}). Recently it was investigated the effect of
the angular momentum perturbation of the Hamiltonian $H_2=\Omega
L_z$ on the energy levels of UCN \cite{arminjon}. Our purpose is
generalize this correction to the case of the gravitating object
(Earth in particular case) which possess also brane parameter.
Denote as the unperturbed non-relativistic stationary state of the
2- spinor (describing UCN) in the field of the rotating
gravitating object in braneworld. Then we have
\begin{eqnarray}\label{Hamil}
H_{4}\psi=i\hbar\frac{Q^*a c}{r^4}\left(1+\frac{2GM}{c^2r}\right)
\frac{\partial\psi}{\partial\phi}=i\hbar\frac{Q^*a c
\sin\theta}{r^3} \left(1+\frac{2GM}{c^2r}\right)\nabla\psi\cdot
e_{\phi} \ ,
\end{eqnarray}
here $\nabla\psi$ is the Laplasian of the spherical coordinates,
which is usually written as:
\begin{eqnarray}
\nabla\psi=\frac{\partial\psi}{\partial
r}e_r+\frac{1}{r}\frac{\partial\psi}{\partial\theta}e_{\theta}
+\frac{1}{r\sin\theta}\frac{\partial\psi}{\partial\phi}e_\phi \ .
\end{eqnarray}
By adopting new Catesian coordinates $x,y,z$ within $e_x\equiv
e_\phi$ and axis $z$ being local vertical, when the stationary
state is assumed to have the form
\begin{eqnarray}
\phi(x)=\phi_\upsilon(z)e^{i(k_1x+k_2y)} \ ,
\end{eqnarray}
one can easily derive from (\ref{Hamil})
\begin{eqnarray}
&&H_4\psi=i\hbar\frac{Q^*a c\sin\theta}{r^3}
\left(1+\frac{2GM}{c^2r}\right)\frac{\partial\psi}{\partial
x}=-\hbar k_1\frac{Q^*a c\sin\theta}{r^3}
\left(1+\frac{2GM}{c^2r}\right)\psi\nonumber\\
&&\qquad\qquad =-mu_1\frac{Q^*a c\sin\theta}{r^3}
\left(1+\frac{2GM}{c^2r}\right)\psi \ ,
\end{eqnarray}
where the following notation has been used
\begin{eqnarray}
u_1\equiv{\bf u}\cdot \mathbf{e}_\phi, \quad {\bf
u}\equiv\frac{\hbar}{m}(k_1\mathbf{e}_x+k_2\mathbf{e}_y) \ .
\end{eqnarray}
Following to the works \cite{arminjon} and \cite{vika} we can
compute modification of the energy level as first-order
perturbation:
\begin{eqnarray}\label{var}
(\delta
E)_{brane}\simeq\langle\psi|H_4\psi\rangle=-mu_1\int\frac{Q^*a c
\sin\theta}{r^3} \left(1+\frac{2GM}{c^2r}\right)|\psi|^2dV \ .
\end{eqnarray}
Assume $r=(R+z)\cos\chi$ (where $\chi$ is the latitude angle) and
$\sin\theta$ to be equal to 1, that is $\theta=\pi/2$. Assuming
now $z\ll{ R}$ one can extend (\ref{var}) as
\begin{eqnarray} (\delta
E')_{brane}\simeq\langle\psi|H_4\psi\rangle=-mu_1\frac{Q^*a c}{{
R}^3\cos^3\chi}+mu_1\frac{3Q^*a c}{{R}^4\cos^3\chi}\int
z|\psi|^2dV \ ,
\end{eqnarray}
and
\begin{eqnarray}
(\delta
E'')_{brane}\simeq\langle\psi|H_4\psi\rangle=-mu_1\frac{2GMQ^*a}
{{R}^4c\cos^4\chi}+mu_1\frac{8GMQ^*a}{{R}^5c\cos^4\chi}\int
z|\psi|^2dV \ ,
\end{eqnarray}
where we have separated equation (\ref{var}) into two parts as
$(\delta E)_{brane}=(\delta E')_{brane}+ (\delta E'')_{brane}$.
Then we remember that $\int z|\psi|^2dV$ is the average value
$\langle z\rangle_n$ of $z$ for the stationary state
$\psi=\psi_n$. For further calculation we need to use formulae for
$\langle z\rangle_n$ from \cite{arminjon}
\begin{eqnarray}
\langle z\rangle_n=\frac{2}{3}\frac{E_n}{mg}\ .
\end{eqnarray}
Now one can easily estimate the relative modification of the
energy level $E_n$ of the neutrons, placed in braneworld.
\begin{eqnarray}
\frac{(\delta E')_{brane}}{E_n}\simeq\frac{2u_1 Q^*a c }{R^4g
\cos^3\chi} \ ,
\quad \quad
\frac{(\delta
E'')_{brane}}{E_n}\simeq\frac{16}{3}\frac{GMQ^*u_1a}{R^5
g\cos^4\chi}=\frac{16}{3}\frac{Q^{*}u_1a}{R^3\cos^4\chi} \ .
\end{eqnarray}
We numerically estimate the obtained modification using the
following parameters for the Earth: $u_1\simeq +10m/s$,
$Q\sim10^9sm$, $\cos\chi\simeq0.71$, $a\simeq3.97 m$,
 $g\simeq10m/s^2$ and $R\simeq6.4\cdot10^{8}sm$, $c\simeq3\cdot10^{8}m/s$
 %$G\simeq6.67\times10^{-11}$
%
\begin{eqnarray}\label{numres}
\frac{(\delta E')_{brane}}{E_n}\simeq4\times10^{-11} \ .
\end{eqnarray}
For the surface of the Earth we can neglect $(\delta E'')_{brane}$
term. %, because of the following condition: $GM/c^2R \ll 1$.
 From the obtained result (\ref{numres}) one can see, that the in
influence of brane parameter will be stronger in the vicinity of
compact gravitating objects with small $R$.
\section{The interference in a Mach-Zehnder-type
interferometer}\label{3rdsec}
\normalsize

Spacetime metric of the rotating black hole in braneworld in
coordinates ${t,r,\theta,\varphi}$ takes form {(see e.g,
\cite{ag05})}
\begin{eqnarray}
&&
ds^2=-\frac{\Delta-a^2\sin^2\theta}{\Sigma}dt^2+\frac{(\Sigma+a^2\sin^2
\theta)^2  -\Delta a^2 }{\Sigma}\sin^2 \theta d\varphi^2
\nonumber\\ &&\qquad\quad +\frac{\Sigma} {\Delta}dr^2 +\Sigma
d\theta^2 -2\frac{\Sigma+a^2\sin^2 \theta-\Delta}{\Sigma}a\sin^2
\theta d\varphi dt\ ,\ \quad\label{metric1}
\end{eqnarray}
where $\Sigma=r^2+a^2\cos^2\theta$, $\Delta=r^2+a^2-2Mr+Q^*$,
$Q^*$ is the bulk tidal charge, $M$ is the total mass and $a$ is
related to the angular momentum of the black hole.

The components of the tetrad frame for the stationary observer for
metric (\ref{metric1}) are
\begin{eqnarray}
\label{zamo_tetrad_1}&&{\bf e}_{\hat
t}^{\mu}=\left(1-\frac{2M}{r}+\frac{Q^{*}}{r^2}\right)^{-\frac{1}{2}}\left(1,0,0,0\right),
\hskip 0.7cm {\bf e}_{\mu}^{\hat
t}=-\left(1-\frac{2M}{r}+\frac{Q^{*}}{r^2}\right)^{\frac{1}{2}}\left(1,0,0,\frac{2Mr-Q^{*}}{r^2}
%-\frac{Q^{*}}{r^2}
\right) \\
\label{zamo tetrad 2} &&{\bf e}_{\hat
r}^{\mu}=\left(1-\frac{2M}{r}+\frac{Q^{*}}{r^2}\right)^{\frac{1}{2}}(0,1,0,0),
\hskip 1.cm {\bf e}_{\mu}^{\hat
r}=\left(1-\frac{2M}{r}+\frac{Q^{*}}{r^2}\right)^{-\frac{1}{2}}(0,1,0,0) \\
\label{zamo tetrad 3} &&{\bf e}_{\hat
\theta}^{\mu}=\frac{1}{r}\left(0,0,1,0\right), \hskip 3.7cm {\bf
e}_{\mu}^{\hat
\theta}=r(0,0,1,0) \\
\label{zamo tetrad 4} &&{\bf e}_{\hat
\varphi}^{\mu}=\frac{1}{r\sin\theta}\left(
%\left
%(1-\frac{2M}{r}+\frac{Q^{*}}{r^2}
%\right
%)^{-\frac{1}{2}}
%\left(
\frac{Q^{*}-{2M}{r}}{r\sqrt{r^2-{2M}{r}+{Q^{*}}}}
%-\frac{2M}{r}
%\right)
,0,0,1\right),
%\hskip 1cm
{\bf e}_{\mu}^{\hat \varphi}=r\sin\theta(0,0,1,1)
\end{eqnarray}
and the acceleration of the Killing trajectories \cite{Valeria} is
\begin{eqnarray}
a_{\mu}=\frac{1}{2}\partial_{\mu}\ln(-g_{00}) \ ,
%=\frac{1}{2}\partial_{\mu}\ln\left(1-\frac{2M}{r}+\frac{{Q}^{*}}{r^{2}}\right)
\end{eqnarray}
and we obtain for nonvanishing component of the acceleration
\begin{equation}
a_{\hat r}=\frac{1}{r}\left(\frac{M}{r}-\frac{Q^{*}}{r^2}\right)
\left(1-\frac{2M}{r}+\frac{Q^{*}}{r^2}\right)^{-\frac{1}{2}} \ .
\end{equation}
%
%\begin{equation}
%a_{\hat\theta}=0
%\end{equation}
%
%\begin{equation}
%a_{\hat\varphi}=0
%\end{equation}
%
 The nonzero components of rotation tensor of the stationary congruence $\chi_{\mu \nu}$ in the
 slowly rotating braneworld are given by
 %
%\begin{equation}
%\chi_{\hat r \hat \theta}=0
%\end{equation}
%
\begin{equation}
\chi_{\hat r \hat
\varphi}=\frac{a\sin\theta}{r^2}\left(1-\frac{2M}{r}+\frac{Q^{*}}{r^2}\right)^{-1}
\left(\frac{M}{r}-\frac{Q^{*}}{r^2}\right)\ ,
\end{equation}
\begin{equation}
\chi_{\hat \theta \hat
\varphi}=\frac{a\cos\theta}{r^2}\left(1-\frac{2M}{r}+\frac{Q^{*}}{r^2}\right)^{-\frac{1}{2}}
\left(\frac{Q^{*}}{r^2}-\frac{2M}{r}\right) \ .
\end{equation}
\begin{center}
\begin{figure}
  \includegraphics[width=0.75\textwidth]{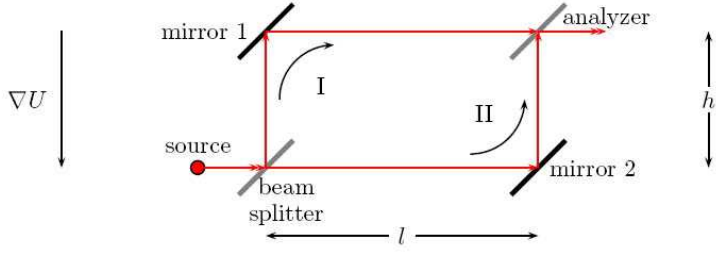}
\caption{{The gradient of $U$ in the interferometer's rest frame.
$h$ and $l$ are the interferometer's height and length.}}
\label{fig:1}       % Give a unique label
\end{figure}
\end{center}

Introducing the $\mathcal{A_{\mu}}$ vector potential of the
electromagnetic field in the Lorentz gauge in simple form
$\mathcal{A}^{\alpha}=C_{1}\xi^{\alpha}_{t}+C_{2}\xi^{\alpha}_{\varphi}$
\cite{Ahmadjon} for the metric (\ref{metrik}) with components.
There the constant $C_{2}=B/2$, where gravitational source is
immersed in the uniform magnetic field $\bf B$ being parallel to
its axis of rotation (properties of black hole immersed in
external magnetic field have been studied in
\cite{Konoplya:2006gg,Konoplya:2006qr}), and the  second constant
$C_{1}=aB$ can be calculated from the asymtotic properties of
spacetime (\ref{metrik}) at the infinity.
\begin{eqnarray}
\mathcal{A}_{t}=-aB\left[1-\left(\frac{2M}{r}-\frac{Q^{*}}{r^2}\right)\left(1-\frac{1}{2}\sin^2\theta\right)\right]
\ ,
\end{eqnarray}
\begin{eqnarray}
\mathcal{A}_{\varphi}=\frac{Br^2}{2}\sin^2\theta \ ,
\end{eqnarray}
we can write the total energy of the particle in the weak field
approximation in the following form:
\begin{eqnarray}
\mathcal{E}=p(\xi)+\mathcal{E}_{pot}=p(\xi)+e_{p}\mathcal{A}_{t} \
,
\end{eqnarray}
where $e_{p}$ is electric charge of the particle. This is
interpreted as total conserved energy consisting of a
gravitationally modified kinetic and rest energy $p(\xi)$, a
modified electrostatic energy $e_{p}\mathcal{A}_{t}$.
  For later use note the measured components of the
  electromagnetic field, which the electric and magnetic fields
  are $E_{\alpha}=F_{\alpha\beta}u^{\beta}$ and
  $B_{\alpha}=(1/2)\eta_{\alpha\beta\mu\nu}F^{\beta\mu}u^{\nu}$,
  where $F_{\alpha\beta}=\mathcal{A}_{\beta,\alpha}-\mathcal{A}_{\alpha,\beta}$ is the
  field tensor,
  $\eta_{\alpha\beta\mu\nu}=\sqrt{-g}\epsilon_{\alpha\beta\mu\nu}$
  is the pseudo-tensorial expression for the Levi-Civita symbol
  $\epsilon_{\alpha\beta\mu\nu}$, $g\equiv det|g_{\alpha\beta}|$.
\begin{eqnarray}
\label{electormagnetic field 1} && B_{\hat r}=-\frac{B}{2} \ ,
\hskip 1cm B_{\hat
\theta}=-\frac{B\sin\theta}{2}\left(1-\frac{2M}{r}+\frac{Q^{*}}{r^2}\right)^{\frac{1}{2}} \ , \\
\label{electormagnetic field 2} && E_{\hat
r}=\frac{aB}{r}\left[2\left(\frac{Q^{*}}{r^2}-\frac{M}{r}\right)+\sin^2\theta\left(\frac{3M}{r}-\frac{2Q^{*}}{r^2}\right)\right],
\hskip 1cm E_{\hat \theta}=0 \ .
\end{eqnarray}
Now with these results we obtain as total phase shift
\cite{Valeria} like in the following form
\begin{eqnarray}
&&\Delta\phi=\mathcal{E}\Sigma\bigg[-\frac{\mathcal{E}}{p_{0}}(\cos{\beta}a_{\hat
r } -\cos{\gamma}\sin{\beta}a_{\hat \theta} -\sin\gamma\sin\beta
a_{\hat \varphi })
%\right.\nonumber\\ &&\left.
 -\frac{1}{p_{0}}(\cos\beta\partial_{\hat
r}\mathcal{E}_{pot} -\cos{\gamma}\sin{\beta}\partial_{\hat
\theta}\mathcal{E}_{pot} \nonumber\\&&\qquad\qquad
-\sin{\gamma}\sin{\beta}\partial_{\hat \varphi}\mathcal{E}_{pot})
+\sin\beta\chi_{\hat\theta \hat
\varphi}+\cos\gamma\cos\beta\chi_{\hat\varphi \hat
r}+\sin\gamma\cos\beta\chi_{\hat r \hat \theta} \bigg]\nonumber\\
 &&\qquad\qquad+e_{p}\Sigma(\sin\beta B_{\hat
r}+\cos\gamma\cos\beta B_{\hat \theta}+\sin\gamma\cos\beta B_{\hat
\varphi}) \ ,
\end{eqnarray}
where $\partial_{\hat \mu}=e_{\hat \mu}^{\nu}\partial_{\nu}$. And
there $\Sigma$ is the area of the interferometer, and $\gamma$ is
the angle of the baseline with respect to ${\bf e}_{\hat
\varphi}$ and $\beta$ the tilt angle.
 Therefore we can independently vary the angles $\beta$ and
 $\gamma$
 %in equatorial plane ($\theta=\pi/2$)
 , where one can extract from phase shift measurements the
 following combinations of terms:
\begin{eqnarray}
&&\Delta\phi\left(\beta=0,\gamma=0\right)=\frac{\Sigma\Lambda\mathcal{L}}{p_0
r\mathcal{W}^{\frac{1}{2}}}(\frac{ap_0}{\mathcal{W}^{\frac{1}{2}}}-\Lambda+e_paB)
-\frac{1}{2}e_pB\Sigma\sin\theta\mathcal{W}\qquad \\
%\end{eqnarray}
%
%\begin{eqnarray}
&&\Delta\phi\left(\beta=\frac{\pi}{2},\gamma=\frac{\pi}{2}\right)=
-\Lambda \frac{\Sigma a\mathcal{K}\mathcal{L}
}{r^2\mathcal{W}^{\frac{1}{2}}}
\cos\theta-\frac{e_pB}{2}\\
%\end{eqnarray}
%
%\begin{eqnarray}
&&\Delta\phi\left(\beta=\frac{\pi}{2},\gamma=0\right)= -\Lambda
\frac{\Sigma a\mathcal{K}
\mathcal{L}}{r^2\mathcal{W}^{\frac{1}{2}}}
\cos\theta-\frac{e_pB}{2}\\
%\end{eqnarray}
%
%\begin{eqnarray}
&&\Delta\phi\left(\beta=0,\gamma=\frac{\pi}{2}\right)=
\frac{\Sigma\Lambda}{p_0r\mathcal{W}^{\frac{1}{2}}}
(e_paB-\Lambda)
\end{eqnarray}
where
\begin{eqnarray}
&& \left(1-\frac{2M}{r}+\frac{Q^*}{r^2}\right)=\mathcal{W},\quad
 \left(\frac{2M}{r}-\frac{Q^*}{r^2}\right)=\mathcal{K}, \quad
 \left(\frac{M}{r}-\frac{Q^{*}}{r^2}\right)=\mathcal{L}\\
 && \Lambda = \left\{p(\xi)
+e_paB\left[1-\mathcal{K}(1-\sin^2\theta/2)\right] \right\}
\end{eqnarray}

Using above obtained results we can estimate upper limit for brane
parameter. Using the results of Earth based atom interferometry
experiments \cite{kasevich} would give us an estimate $Q^*\leq
10^{8}$ cm$^2$

\section{The Sagnac effect}\label{4thsec}

\normalsize As it was showed that the Sagnac effect for
counter-propagating on a round trip in a interferometer rotating
in a flat space-time, may be obtained by a formal analogy with the
Aharonov-Bohm effect. Here we study the interference process of
matter or light beams in the spacetime of slowly rotating compact
gravitating (say Earth) in braneworld in terms of the
Aharonov-Bohm effect \cite{ruggiero1}. The phase shift
\begin{eqnarray}\label{phaseshift}
&& \Delta\phi=\frac{2m \rho_0}{c\hbar}\oint_C {\bf A}_G\cdot dx ,
\end{eqnarray}
detected by a uniformly rotating interferometer, and the time
difference between the propagation times of the co-rotating and
counter-rotating beams will be equal to
\begin{eqnarray}\label{timedelay}
&& \Delta T=\frac{2\rho_0}{c^3}\oint_C {\bf A}_G\cdot dx .
\end{eqnarray}

In the expressions (\ref{phaseshift}) -- (\ref{timedelay}) $m$
indicates the mass (or the energy) of the particle of the
interfering beams , $A_G$ is the gravito-magnetic vector
potential, which is obtained from the expression
\begin{eqnarray}\label{gravito}
{\bf A}^{G}_{i}\equiv c^2\frac{\rho_i}{\rho_0} ,
\end{eqnarray}
and $\rho(x)$ is the unit four-velocity of particles:
\begin{equation}
\rho^\alpha\equiv\left\{\frac{1}{\sqrt{-g_{00}}},0,0,0\}\right\},
\quad \rho_\alpha\equiv\{-\sqrt{-g_{00}},g_{i0}\rho^0\}.
\end{equation}
where $g_{\alpha\beta}$ is the components of the metric
(\ref{metric})
In equatorial plane $(\theta=\pi/2)$ we apply the coordinate
transformation $\phi\rightarrow\phi+\Omega t$ to the metric
(\ref{metric}), where $\Omega$ is the angular velocity of the
gravitating object, we obtain
\begin{equation}\label{metric}
ds^{2}=-(A^{2}-r^{2}\Omega^{2}+2\widetilde{\omega}
r^{2}\Omega)dt^{2}+H^{2}d
r^2+r^{2}d\phi^{2}+2r^{2}(\Omega-\widetilde{\omega})dtd\phi,
\end{equation}
for simplicity we expand the $H^2$ in the terms of $1/r$, then one
can easily obtain the following expression:
\begin{equation}
H^{2}\simeq 1+\frac{2M}{r}-\frac{Q^{*}}{r^{2}}
\end{equation}
From this equation one can see, that the unit vector field
$\rho(x)$ along the trajectories $r=R=\rm{const}$ will be
$$
\rho_{0}=-(\rho^{0})^{-1}
$$
\begin{equation}
\rho_{\phi}=r^{2}(\Omega-\widetilde{\omega})\rho^{0},
\end{equation}
where we have used the following notation
\begin{equation}
\rho^{0}=\left[1-\frac{2M}{r}+\frac{Q^{*}}{r^2}-r^{2}
\Omega(\Omega-\widetilde{\omega})\right]^{-1/2} \ .
\end{equation}
Now, inserting the components of $\rho(x)$ into the equation
(\ref{gravito}) one obtain
\begin{equation}
{\bf A}^{G}_{\phi}=-r^{2}(\Omega-\widetilde{\omega})(\rho^{0})^2
\end{equation}

Integration vector potential, as it is shown in equation
(\ref{phaseshift}) and (\ref{timedelay}), one can get the
following expressions for $\Delta\phi$ and $\Delta T$ (here we
returned to the physical units):
\begin{equation}
\Delta\phi =\frac{4\pi
m}{\hbar}r^{2}(\Omega-\widetilde{\omega})\rho^{0}
\end{equation}
\begin{equation}\label{ABDelta T}
\Delta T=\frac{4\pi}{c^{2}}r^{2}(\Omega-\widetilde{\omega})
\rho^{0}
\end{equation}

Following to the paper \cite{ruggiero1} one can find a critical
angular velocity $\bar{\Omega}$
\begin{equation}
\bar{\Omega}=\widetilde{\omega}=\omega\left(1-\frac{Q^{*}}{2r
m}\right)
\end{equation}
which corresponds to zero time delay $\Delta T=0$. $\bar{\Omega}$
is the angular velocity of zero angular momentum observers (ZAMO).
As we remember that brane parameter is a negative, then one can
see that parameter $\bar{\Omega}$ increases in braneworld.

\section{Conclusion}\label{conclusion}

In the present paper we have considered quantum interference
effects in  spacetime of black hole on braneworld and found that
the presence of brane parameter in the metric can have influence
on different quantum phenomena. Namely, we have obtained the phase
shift and time delay in Sagnac effect can be affected by brane
parameter. Then, we have found an expression for the phase shift
in a neutron interferometer due to existence of tidal charge and
studied the feasibility of its detection with the help of
"figure-eight" interferometer. We have also investigated the
application of obtained results to the calculation of energy
levels of UCN and found modifications to be rather small for the
Earth, but maybe more relevant for compact astrophysical objects.
The result shows that the phase shift for a Mach-–Zehnder
interferometer in spacetime of gravitational object on braneworld
is influenced by brane parameter. Obtained results can be further
used in experiments to detect the interference effects related to
the phenomena of braneworld. Recently authors of the paper
\cite{bohmer} got estimation upper limit for brane parameter as
$Q^*\leq 10^8 \rm{cm}^2$ from classical Solar system tests. In the
paper \cite{sepangi} also made some good estimations for brane
parameter from effects around planets in Solar system. Here we
estimated upper limit for brane parameter as $Q^*\leq 10^{7}$
cm$^2$ using Earth based experiments \cite{colella}.

\section*{Acknowledgments}

The work was supported by the UzFFR (projects 5-08 and 29-08) and
projects FA-F2-F079 and FA-F2-F061 of the UzAS. This work is
partially supported by the ICTP through the OEA-PRJ-29 project.
Authors gratefully thank Viktoriya Giryanskaya, Ahmadjon
Abdujabbarov and Bobomurat Ahmedov for useful discussions and
invaluable help in the process of preparation of the work.

% references


\newpage
\begin{thebibliography}{99}

\bibitem{randal}{L. Randall and R. Sundrum, Phys. Rev. Lett.
{\bf 83}, 3370; 4690 (1999).}

\bibitem{dadhich}{N.K. Dadhich, R. Maartens, P.Papodopoulos, and
V.Rezania, Phys. Lett. B \textbf{487}, 1 (2000).}

\bibitem{maartens}{R. Maartens, Living Rew. Relativity 7, 7(2004).}

\bibitem{stuklik} Z. Stuchlik and A. Kotrlov\'{a}, Gen. Rel.
Grav., \textbf{41}, 1305 (2009).

\bibitem{stuklik2} A. Kotrlov\'{a}, Z. Stuchlik, and G. T\"{o}r\"{o}k, Class.
Quantum Grav., \textbf{25}, 225016 (2008).

\bibitem{overhauser}{A.W. Overhauser and R.Colella, Phys. Rev.
Lett.  \textbf{33}, 1237 (1974).}

\bibitem{colella} {R.Colella, A.W. Over Hauser, and S.A. Werner,
Phys. Rev. Lett. \textbf{34}, 1472 (1975).}

\bibitem{page}{L.A. Page, Phys. Rev. Lett. \textbf{35}, 543 (1975).}

\bibitem{werner}{S.A. Werner, J.L. Staudenmann, and R.Colella,
Phys. Rev. Lett. \textbf{42}, 1103 (1979).}

\bibitem{mashhoon} {B.Mashhoon, F.W. Hehl, and D.S. Theiss, Gen.
Rel. Grav. \textbf{16}, 711 (1984).}

\bibitem{kuroiwa}{J. Kuroiwa, M.Kasai, and T. Futamase, Phys.
Lett. A \textbf{182}, 330 (1993).}

\bibitem{G.rizzi}{G. Rizzi and M.L. Ruggiero,  gr-qc/0305084
(2004).}

\bibitem{rizzi}{G.Rizzi and M.L. Ruggiero, Gen.Rel. Grav.
\textbf{35}, 1743 (2003).}

\bibitem{ruggiero}{M.L. Ruggiero, Gen. Rel. Grav. \textbf{37},
1845 (2005).}

\bibitem{nesvizhevsky} {V. V. Nesvizhevsky et. al., Phys.
Rev. D \textbf{67}, 102002 (2003).}

%\bibitem{arifov73} {L.Ya. Arifov, Doklady Akad. Nauk SSSR
%\textbf{210}, 1320 (1973), in Russian.}


\bibitem{arminjon}{M. Arminjon, Phys. Let. A \textbf{372}, 2196
(2008).}

\bibitem{vika}{V.S. Morozova and B.J. Ahmedov, Int. J. Mod. Phys. D
\textbf{18}, 107 (2009).}
%
\bibitem{ag05}
{A. N. Aliev and A. E. G\"{u}mr\"{u}k\c{c}\"{u}o\v{g}lu, Phys.
Rev. D {\bf{71}}, 104027 (2005).}
%
\bibitem{Valeria}{V. Kagramanova, J. Kuntz, and C.
L\"{a}mmerzahl, Class. Quantum Grav. \textbf{25}, 105023 (2008).}

\bibitem{Ahmadjon}{A. A. Abdujabbarov, B. J. Ahmedov, and V. G.
Kagramanova, Gen. Rel. Grav. \textbf{40}, 2515(2008). }

\bibitem{Konoplya:2006gg}  {R.~A.~Konoplya,  Phys.\ Lett.\  B {\bf
644}, 219 (2007), [arXiv:gr-qc/0608066].}

\bibitem{Konoplya:2006qr} R.~A.~Konoplya, Phys.\ Rev.\  D {\bf
74}, 124015 (2006), [arXiv:gr-qc/0610082].

\bibitem{kasevich}{ S. Dimopoulos, P.W. Graham, J.M. Hogan, and M.E. Kasevich.
Phys. Rev. D, \textbf{78}, 042003 (2008).}

\bibitem{ruggiero1}{M.L. Ruggiero, Gen. Rel.Grav. \textbf{37},
1845 (2005).}

\bibitem{bohmer} C.G. B\"{o}hmer, T. Harko, and F. S. N. Lobo, Classical Quantum
Gravity  {\bf 25}, 045015 (2008).

\bibitem{sepangi}{S. Jalalzadeh., M. Mehrnia, and H. R. Sepangi,
arXiv:0906.4404v1 [gr-qc] (2009).}


%\bibitem{mak1}{V.S. Morozova, B.J. Ahmedov, and V.G. Kagramanova,
%Astrophys. J., \textbf{684}, 1359 (2008).}

\end{thebibliography}
\end{document}